\documentclass[a4paper,english,reprint, aps,pre]{revtex4-1}
\usepackage[T1]{fontenc}
\usepackage[latin9]{inputenc}
\usepackage{array}
\usepackage{verbatim}
\usepackage{dvipost}
\usepackage{amsmath}
\usepackage{amssymb}
\usepackage{graphicx}

\makeatletter

\providecommand{\tabularnewline}{\\}
\dvipostlayout
\dvipost{osstart color push Red}
\dvipost{osend color pop}
\dvipost{cbstart color push Blue}
\dvipost{cbend color pop}

\newcommand{\ket}[1]{\left| #1 \right>} 
\newcommand{\bra}[1]{\left< #1 \right|} 
\PassOptionsToPackage{caption=false}{subfig}

\@ifundefined{showcaptionsetup}{}{%
 \PassOptionsToPackage{caption=false}{subfig}}
\usepackage{subfig}
\makeatother

\usepackage{babel}
\begin{document}

\title{Dissipation production in a closed two-level quantum system as a
test of the obversibility of the dynamics}

\author{Claudia L. Clarke and Ian J. Ford}

\affiliation{Department of Physics and Astronomy and London Centre for Nanotechnology,
University College London, Gower Street, London WC1E 6BT, U.K.}
\begin{abstract}
Irreversible behaviour is traditionally associated with open stochastic
dynamical systems, but an asymmetry in the probabilistic specification
of a closed deterministic system can similarly lead to a disparity
between the likelihoods of a particular forward and corresponding
backward behaviour starting from a specified time. Such a comparison
is a test of a property denoted \emph{obversibility}, which may be
quantified in terms of \emph{dissipation production} as a measure
of irreversibility. We here discuss the procedure needed to evaluate
dissipation production in a simple, deterministic two-level quantum
system described by a statistical ensemble of state vectors and then
provide numerical results for illustrative situations. We consider
cases that both do and do not fulfill an Evans-Searles Fluctuation
Theorem for the dissipation production, and identify conditions for
which the system will display time-asymmetric average behaviour as
it evolves.
\end{abstract}



\maketitle

\section{Introduction}

Irreversible behaviour, which cannot be reversed or undone, is ubiquitous
in everyday life. Each time an ice cube melts, reaching room temperature
as it does so, we observe macroscopic irreversibility. Other examples
include system mixing and spreading. Such behaviour is traditionally
characterised by a monotonic rise in entropy, in accordance with the
second law of thermodynamics \citep{kreuzer1981nonequilibrium}. This
allows past to be distinguished from present, identifying the future
as the direction in which entropy increases, leading to the emergence
of an \emph{arrow of (development in) time} \citep{lebowitz1993boltzmann}\emph{. }

However, given that the Newtonian laws governing macroscopic motion
are time reversal symmetric, we still seek, as did Boltzmann at the
advent of the study of thermodynamics, a satisfactory understanding
of the emergence of macroscopic irreversibility. One proposed explanation
is known as the Past Hypothesis. This posits that entropy increases
globally because the universe started from a state of low entropy
so it is likely that all conceivable evolutions lead to its increase
\citep{albert2009time}. The irreversibility, and the evolution of
its measure, entropy, ultimately depend on the initial conditions
taken by a system. The persistent impact of initial conditions may
have philosophical, as well as physical consequences, which can be
explored \citep{albert2009time}.

{} 

Closed, deterministically evolving systems are mechanically reversible,
meaning that a protocol of manipulation can be followed which returns
an evolved state to its initial state. A forward trajectory can then
always be followed by a backward trajectory. Entropy production \citep{tolman1948irreversible}
is primarily intended for use as a measure of irreversibility for
\emph{open} systems, and is intimately connected to the development
of uncertainty and loss of information \citep{eckmann1985ergodic}
brought about by environmental interactions or the process of measurement
in quantum systems. It explores a failure of mechanical reversibility,
a distinction between the probabilities of forward and subsequent
backward evolution, and hence acts as a measure of irreversibility.
Progress has been made in understanding entropy production in open
classical dynamical systems, as well as in studying the average \citep{deffner2011nonequilibrium}
and single realisation quantum entropy production in quantum systems
in the weak coupling limit \citep{horowitz2013entropy,elouard2017probing}.
However, for systems that are closed and dynamically deterministic,
such a measure of irreversibility is not available and some other
approach needs to be employed in order to quantify the irreversible
behaviour that can emerge in many situations.

We therefore consider an alternative measure that, like entropy production,
derives from a difference in the likelihood of forward and backward
trajectories, but under deterministic, mechanically reversible dynamics.
To distinguish this property from the entropy measure that tests the
reversibility, we call this a test of the \emph{obversibility} of
the dynamics, a concept that has previously been investigated in classical
situations \citep{Ford,evans2002fluc}.We now consider it for the
first time in a quantum context. We provide a definition of obversibility
and quantify its measure as \emph{dissipation production}.The nomenclature
chosen is intended to evoke parallels with the established related
quantifier of irreversibility, \emph{entropy production}, and follows
from an earlier name for the measure used in previous literature:
the time-integrated modified dissipation function \citep{Ford}, which
is a generalisation of the time-integrated dissipation function coined
by Evans \citep{evans1993probability}. In a study of dissipation
production in a purely deterministic system without environment, it
is a property unrelated to traditional notions of the dissipation
of heat. Instead, dissipation production is associated with the observation
of unexpected behaviour, akin to microscopic violations of the second
law. The entropy production is more specifically associated with the
failure of a subsequent \emph{negation} of behaviour. However, if
the system were coupled to an environment, then there would be traditional
heat dissipation.

We evaluate dissipation production and explore some of its properties
in the simplest case of a closed two-level quantum system evolving
without measurement, for which entropy production is not an applicable
irreversibility measure. We demonstrate that dissipation production
may be computed for individual realisations of the system dynamics,
that its average over all possible realisations is never negative,
and that in certain situations its probability density function (pdf)
can satisfy a symmetry known as the Evans-Searles Fluctuation Theorem
(ESFT). When the ESFT is \emph{not} satisfied, the average dissipation
production of the system as it evolves into the future differs from
its average evolution into the past.

\section{Methods\label{sec:Methods}}

Classically, the forward trajectory taken by a system is simply the
path it follows through phase space from an initial configuration
defined by a point in phase space, $\Gamma_{A}$, to a final configuration,
$\Gamma_{B}$. Translating to a quantum setting, these might correspond
to initial and final wavefunctions $\psi_{A}$ and $\psi_{B}$, which
are defined by a collection of complex probability amplitudes $\widetilde{\Gamma}_{A}$
and $\widetilde{\Gamma}_{B}$ respectively, according to some basis
set. In order to define a backward trajectory, we employ an inversion
operator $M^{T}$ which has the effect of transforming the final configuration
reached along the forward trajectory into an appropriate starting
point from which the backward trajectory can proceed under the reversed
dynamics.  In a classical situation, this transformation is \emph{velocity
inversion}, $v\rightarrow M^{T}v=-v$ \citep{Ford}. However, the
equivalent reversal in the quantum case is conjugation of the wavefunction,
$\psi\rightarrow M^{T}\psi=\psi^{*}$, as this sets up conditions
for a time reversed solution to the Schr{\"o}dinger equation \citep{haake2013quantum}.
We associate a set of amplitudes $\widetilde{\Gamma}^{*}$ with each
$\psi^{*}$ and define a mapping $M^{T}$ such that $M^{T}\widetilde{\Gamma}=\widetilde{\Gamma}^{*}$.

We make the distinction between the quantum uncertainty associated
with measurement outcome, and the classical uncertainty associated
with the choice of state that embodies particular probabilities for
given measurement outcomes. As we are considering deterministic dynamics,
we only consider the classical uncertainty and postulate that the
pdf encoding the probability of finding certain quantum states has
ontological reality, even though it is often combined with the quantum
measurement uncertainty when using a density matrix. Nevertheless,
though the pdf for states may be inaccessible, it is a meaningful
concept, particularly if situations are considered in which we have
control over the generation of initial states, and can ascribe particular
probabilities to particular quantum states; pdfs can be used to construct
density matrices, even if they cannot subsequently be extracted from
one. Therefore, we proceed by assigning classical probabilities to
the sets of numbers $\widetilde{\Gamma}$ that define quantum states.

Thus, starting from $\Gamma_{B}$ or in the quantum case $\widetilde{\Gamma}_{B}$,
the application of an inversion operator $M^{T}$ followed by reversed
dynamics over a further time period $t$ returns a system to the inverted
form of original state if the dynamics are mechanically reversible.
However, if the dynamics are indeterminate, with the evolution represented
in terms of probabilities, the degree of failure of reversibility
can be quantified with the \emph{stochastic entropy production,} $\Delta s_{t}$,
a measure of the increase in uncertainty brought about by the dynamics
of a system's evolution and defined as

\begin{equation}
\Delta s_{t}=\mbox{ln}\frac{f(\widetilde{\Gamma}_{A},0)d\widetilde{\Gamma}_{A}\,T(\widetilde{\Gamma}_{A}\rightarrow\widetilde{\Gamma}_{B})}{f(\widetilde{\Gamma}_{B},t)d\widetilde{\Gamma}_{B}\,P_{I}(\widetilde{\Gamma}_{B}\rightarrow\widetilde{\Gamma}_{B}^{*})\,T(\widetilde{\Gamma}_{B}^{*}\rightarrow\widetilde{\Gamma}_{A}^{*})},\label{eq:1-1}
\end{equation}
in notation suitable for a quantum setting. Replacing $\widetilde{\Gamma}_{i}$
with $\Gamma_{i}$ gives the appropriate expression for the classical
stochastic entropy production. In Eq. (\ref{eq:1-1}), $f(\widetilde{\Gamma},\tau)$
is the pdf of the configuration of quantum probability amplitudes,
$\widetilde{\Gamma}$, at time $\tau$, such that $f(\widetilde{\Gamma},\tau)d\widetilde{\Gamma}$
is the probability that the configuration lies in the region $d\widetilde{\Gamma}$
about $\widetilde{\Gamma}$. $f(\widetilde{\Gamma},\tau)$ in this
context is essentially the\emph{ classical} probability density of
selecting a certain quantum state vector from an ensemble of possibilities.
Each quantum state vector is itself associated with probabilities
for finding one outcome or another upon measurement, but as stated
earlier we restrict ourselves here to considering deterministic quantum
evolution \emph{without} measurement. We are hence considering the
irreversibility associated only with the evolution of a deterministic,
unmonitored quantum system. In particular, we provide results for
closed systems, which is adequate to show that given particular conditions
on the initial pdf and subsequent evolution, even this simple, restricted
system will show time asymmetric average behaviour, distinct from
the time asymmetric average behaviour assessed through entropy production. 

In Eq. (\ref{eq:1-1}) $T(\widetilde{\Gamma}\rightarrow\widetilde{\Gamma}^{\prime})$
is the probability for a transition from $\widetilde{\Gamma}$ to
$\widetilde{\Gamma}^{\prime}$ according to the dynamics in a time
interval of length $t$. The probability for the inversion $P_{I}(\widetilde{\Gamma}_{B}\rightarrow\widetilde{\Gamma}_{B}^{*})$
in the denominator of Eq. (\ref{eq:1-1}) might be omitted since it
is unity, though its presence makes more apparent the precise nature
of the two processes that are being compared. The inversion operation
is taken to act instantaneously. The idea of Eq. (\ref{eq:1-1}) is
to compare the probability of a forward path from $\widetilde{\Gamma}_{A}$
to $\widetilde{\Gamma}_{B}$, in a time interval of length $t$, with
the probability of \emph{subsequently} starting from a configuration
$\widetilde{\Gamma}_{B}$ at time $t$, inverting it and having it
return to configuration $\widetilde{\Gamma}_{A}^{*}$ after dynamical
evolution for a further time $t$. For stochastic dynamics, the ratio
of initial to final  increments $d\widetilde{\Gamma}_{A}$/$d\widetilde{\Gamma}_{B}$
is unity and can be omitted.

For closed systems with deterministic dynamics, $\Delta s_{t}$ vanishes
since the transition probabilities $T$ are replaced by deterministic
mappings of the state, taken with unit probability. Equivalently,
the evolution of $\widetilde{\Gamma}_{A}$ to $\widetilde{\Gamma}_{B}$
might be represented by the operation $S_{t}$, the backward trajectory
by $S_{t}^{*}$ and, including the inversions, the reversibility of
the dynamics corresponds to $M^{T}S_{t}^{*}M^{T}S_{t}\widetilde{\Gamma}_{A}=\widetilde{\Gamma}_{A}$\emph{.}
By conservation of probability we have $f(\widetilde{\Gamma}_{B},t)d\widetilde{\Gamma}_{B}=f(\widetilde{\Gamma}_{A},0)d\widetilde{\Gamma}_{A}$
and hence $\Delta s_{t}=0$. There is no entropy production since
there is no change in the uncertainty of the state brought about by
the dynamics. 

So for closed, deterministic systems we therefore need a different
quantity with which to measure irreversibility. A suitable quantity
called the \emph{dissipation production} has been employed in classical
situations \citep{Ford}, developing earlier work by Evans \citep{evans1993probability}.
Rather than comparing the likelihoods of the forward and (subsequent)
backward trajectories to quantify irreversibility, as in the case
of entropy production, dissipation production compares the likelihoods
of the forward and \emph{obverse} trajectories. In the quantum situation,
the \emph{obverse trajectory} takes the inverted final configuration,
$\widetilde{\Gamma}_{B}^{*}$, via the reversed dynamics, $S_{t}^{*}$,
to the inverted initial configuration $\widetilde{\Gamma}_{A}^{*}$,
but starting the reverse evolution at time \emph{zero}, rather than
at time $t$, which distinguishes it from the backward trajectory,
which is the subsequent reversal of the previous forward trajectory.
We assume that the probability density at $t=0$ is such that all
possible final configurations are accessible. By comparing the likelihood
of a system being in $\widetilde{\Gamma}_{A}$ at $t=0$ with the
likelihood that it begins in $\widetilde{\Gamma}_{B}$, we can quantify
the failure of \emph{obversibility}, a counterpart to the reversibility
that is tested by the entropy production. Such failure is necessarily
and sufficiently a consequence of the properties of the initial probability
density over the configuration space, rather than of the dynamics.

The dissipation production is defined as $\omega_{t}=\mbox{ln}[f(\widetilde{\Gamma}_{A},0)d\widetilde{\Gamma}_{A}/f(\widetilde{\Gamma}_{B},0)d\widetilde{\Gamma}_{B}]$,
where $\widetilde{\Gamma}_{A}$ and $\widetilde{\Gamma}_{B}$ are
related by the mapping $\widetilde{\Gamma}_{B}=S_{t}\widetilde{\Gamma}_{A}$.
To make a more exact parallel with the definition of stochastic entropy
production, dissipation production could also be written as 
\begin{equation}
\omega_{t}=\mbox{ln}\frac{f(\widetilde{\Gamma}_{A},0)d\widetilde{\Gamma}_{A}\,\tilde{T}(\widetilde{\Gamma}_{A}\rightarrow\widetilde{\Gamma}_{B})}{f(\widetilde{\Gamma}_{B},0)d\widetilde{\Gamma}_{B}\,P_{I}(\widetilde{\Gamma}_{B}\rightarrow\widetilde{\Gamma}_{B}^{*})\,\tilde{T}(\widetilde{\Gamma}_{B}^{*}\rightarrow\widetilde{\Gamma}_{A}^{*})},\label{eq:fulldisprod}
\end{equation}
where the appropriate transition probabilities $\tilde{T}$ are unity
since we are considering deterministic dynamics under which $\widetilde{\Gamma}_{A}$
inevitably evolves into $\widetilde{\Gamma}_{B}$, and $\widetilde{\Gamma}_{B}^{*}$
into $\widetilde{\Gamma}_{A}^{*}$ (the latter under reversed dynamics).
Hence $\tilde{T}$ can be omitted, together with the inversion probability
$P_{I}(\widetilde{\Gamma}_{B}\rightarrow\widetilde{\Gamma}_{B}^{*})$
in the denominator. Unlike entropy production, which compares the
likelihood of a system evolving forward and then \emph{subsequently}
evolving backward, dissipation production is evaluated using only
the pdf at an initial time $t=0$. It is a comparison of the probabilities
of \emph{one event or another}. The evolution under $S_{t}$ taking
$\widetilde{\Gamma}_{A}$ to $\widetilde{\Gamma}_{B}$ and the evolution
under $S_{t}^{*}$ taking $\widetilde{\Gamma}_{B}^{*}$ to $\widetilde{\Gamma}_{A}^{*}$
both take place in a time interval \emph{t. }A comparison of the tests
for reversibility and obversibility is made in Table \ref{tab:disprodentprodcomp}. 

\begin{table}
\begin{tabular}{|>{\centering}p{0.2\columnwidth}|>{\centering}p{0.35\columnwidth}|>{\centering}p{0.27\columnwidth}|}
\hline 
 & Compare likelihood of forward \emph{then} reverse paths  & Compare likelihood of forward \emph{or} reverse paths \tabularnewline
\hline 
\hline 
concept tested & reversibility & obversibility\tabularnewline
\hline 
quantifying property  & entropy production

$\Delta s_{t}=\mbox{ln}\frac{f(\widetilde{\Gamma}_{A},0)T(\widetilde{\Gamma}_{A}\rightarrow\widetilde{\Gamma}_{B})}{f(\widetilde{\Gamma}_{B},t)T(\widetilde{\Gamma}_{B}^{*}\rightarrow\widetilde{\Gamma}_{A}^{*})}$ & dissipation production 

$\omega_{t}=\mbox{ln}\frac{f(\widetilde{\Gamma}_{A},0)}{f(\widetilde{\Gamma}_{B},0)}$\tabularnewline
 &  & \tabularnewline
\hline 
\end{tabular}

\caption{Comparison of measures of irreversibility. For simplicity, and a more
compact expression, we assume that the deterministic dynamics conserves
increments in configuration space. $f(\widetilde{\Gamma},t)$ is the
pdf describing an ensemble of sets of probability amplitudes that
define the state vectors and $T$ is a transition probability under
stochastic dynamics. }
\label{tab:disprodentprodcomp}
\end{table}

Dissipation production is clearly zero when states $\widetilde{\Gamma}_{A}$
and $\widetilde{\Gamma}_{B}$ are equally likely to be selected from
the ensemble at $t=0$, in which case we say the evolution is obversible.
If they are not equally likely, but the evolution interval is short,
such that $\widetilde{\Gamma}_{B}$ lies close to $\widetilde{\Gamma}_{A}$,
the dissipation production will likewise be small. However, $\widetilde{\Gamma}_{B}^{*}$
will typically be distant from $\widetilde{\Gamma}_{A}$, and this
is one reason it is important to define dissipation production in
terms of a ratio of $f(\widetilde{\Gamma}_{A},0)$ to $f(\widetilde{\Gamma}_{B},0)$
rather than to $f(\widetilde{\Gamma}_{B}^{*},0)$. 

\subsection{Bloch Sphere Representation\label{subsec:Bloch-Sphere-Representation}}

In the two-level quantum case we shall be considering, a general $\psi$
can be written (now using ket notation) in the form

\begin{equation}
\ket\psi=\cos(\theta/2)\ket{0}+e^{i\phi}\sin(\theta/2)\ket{1},
\end{equation}
where $\cos(\theta/2)$ and $e^{i\phi}\sin(\theta/2)$ are the amplitudes
forming the associated configuration $\widetilde{\Gamma}$. We can
then use Cartesian co-ordinates defined as 

\begin{eqnarray}
x & = & \sin\theta\cos\phi\ ,\ y=\sin\theta\sin\phi\ ,\ z=\cos\theta
\end{eqnarray}
to represent configurations as points on a \emph{Bloch sphere }\citep{nielsen2002quantum}\emph{.}
Trajectories are then paths from an initial point on a Bloch sphere
to the point representing the evolved configuration. 

Unitary evolutions of a two-level quantum system, representing the
mapping $S_{t}$, can be illustrated as continuous paths on the surface
of a Bloch sphere, by analogy with a classical trajectory through
coordinate phase space. Without loss of generality we consider the
mapping from an initial to a final state to be a rotation with unitary

\begin{equation}
S_{t}(\mathbf{\hat{n}},\alpha_{{\rm rot}}(t))=\mathbb{I}\,\mbox{cos}(\alpha_{{\rm rot}}/2)-i\mathbf{\hat{n}}\cdot\hat{\boldsymbol{\sigma}}\,\mbox{sin}(\alpha_{rot}/2),
\end{equation}
where $\mathbf{\hat{n}}$ is the (normalised) axis of rotation, $\alpha_{{\rm rot}}$
is the angle of rotation and $\hat{\boldsymbol{\sigma}}$ is the vector
of Pauli matrices. The angle of rotation is a function of the duration
of the evolution. For a time-independent Hamiltonian, the trajectories
are precisely rotations on the Bloch sphere, with elapsed time proportional
to angle rotated, but the initial to final mapping (though not the
intervening behaviour) can also represent the effect of a Hamiltonian
that is time-dependent. As we are considering only deterministic evolution,
no measurement stage is involved, as this would introduce uncertainty
of outcome into the dynamics. 

In the Bloch sphere representation, the inversion operation, complex
conjugation, is the transformation $\phi\rightarrow-\phi$. Furthermore,
the dynamics conserve areas on the surface of the Bloch sphere, in
the sense that a patch of size $d\widetilde{\Gamma}_{A}$ is mapped
to an equal size patch $d\widetilde{\Gamma}_{B}$. We shall therefore
be able to omit these increments in the definition of the dissipation
production for this case.

The choice of basis used to specify the ensemble of quantum state
vectors, is arbitrary, determining only which states are found at
the poles of the Bloch sphere.%
As unitary transformations are rotations on the Bloch sphere, a change
of basis changes the axes of the Bloch sphere, defined by the basis
vectors, but not any feature such as the pdf displayed upon it. As
the dissipation production is defined in terms of this pdf the particular
choice of basis will not affect the values of \emph{$\omega_{t}$}. 

\begin{figure}
\includegraphics[width=1\columnwidth]{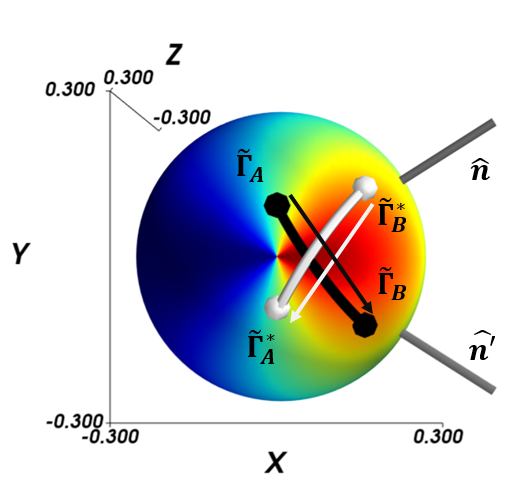}\label{fig:sphere}

\caption{Forward ($\widetilde{\Gamma}_{A}$ to $\widetilde{\Gamma}_{B}$) and
obverse ($\widetilde{\Gamma}_{B}^{*}$ to $\widetilde{\Gamma}_{A}^{*}$)
trajectories shown on the Bloch sphere, the colour of which denotes
the probability density of the initial configuration. The forward
and obverse trajectories are produced by rotations about the $\hat{\mathbf{n}}$
and $\hat{\mathbf{n}}^{\prime}$ axes, respectively. }
\end{figure}

\subsection{Mathematical Properties of Dissipation Production}

\subsubsection{Non-negativity of Mean Dissipation Production}

It can be shown that the mean of the dissipation production will never
be negative, even under reversible dynamics. This property indicates
that, just like entropy production, dissipation production satisfies
a second law-like relation; its \emph{average} behaviour is to increase
as time passes \citep{reid2012communication}. 

To prove the non-negativity of the mean dissipation production, we
start with the expression for the mean, which is

\begin{equation}
\langle\omega_{t}\rangle=\int f(\widetilde{\Gamma},0)\mbox{ln}\frac{f(\widetilde{\Gamma},0)}{f(\widetilde{\Gamma}_{t},0)}d\widetilde{\Gamma},
\end{equation}
where $\widetilde{\Gamma}$ and $\widetilde{\Gamma}_{t}=S_{t}\widetilde{\Gamma}$
are configurations and $f(\widetilde{\Gamma},\tau)$ is the pdf over
configurations at time $\tau$. Note that this expression takes the
form of a Kullback-Leibler divergence \citep{kullback1951information},
or relative entropy, between the initial and shifted pdfs, and a Kullback-Leibler
divergence is never negative. 

Equivalently, it can be shown that the mean dissipation production
is a non-negative quantity by considering the average of its negative
exponential:

\begin{equation}
\langle e^{-\omega_{t}}\rangle=\int f(\widetilde{\Gamma},0)\frac{f(\widetilde{\Gamma}_{t},0)}{f(\widetilde{\Gamma},0)}d\widetilde{\Gamma}=\int f(\widetilde{\Gamma}_{t},0)d\widetilde{\Gamma}.
\end{equation}
Since $f(\widetilde{\Gamma},0)$ is a normalised pdf, and the transformation
$\widetilde{\Gamma}\rightarrow\widetilde{\Gamma}_{t}$ has a Jacobian
of unity, we can write $\langle e^{-\omega_{t}}\rangle=1$. Using
the expansion of $e^{z}$ to establish that $e^{-z}\geq1-z,$ $z\in\mathbb{R}$,
it follows that $\langle e^{-\omega_{t}}\rangle\geq1-\langle\omega_{t}\rangle$
such that $1\geq1-\langle\omega_{t}\rangle$, allowing us to conclude
that $\langle\omega_{t}\rangle\geq0$. It should be noted that this
emerges for both positive and negative $t$, namely evolution into
the future and into the past relative to the starting condition.

\subsubsection{Fluctuation Relation}

A \textit{fluctuation relation} \citep{evans1993probability,evans2002fluc,harris2007fluctuation}
quantifies the extent to which a property such as entropy production
evolves in a direction counter to that dictated by the second law
of thermodynamics. The implication of such a relation is that fluctuations
that `break' the second law are exponentially unlikely and are never
apparent on a macroscopic scale.A negative value for the dissipation
production $\omega_{t}$ indicates behaviour which violates a second
law-like relation.

Entropy production is known to obey a number of fluctuation relations
\citep{PhysRevE.60.2721}. Similarly, the dissipation production $\omega_{t}$
can satisfy a result known as the Evans-Searles Fluctuation Theorem
(ESFT) in certain situations, which we now explore. The requirements
\citep{evans2002fluc,Ford} are that the probabilities of two starting
points of the evolution, related by a mapping $M^{R}$, should be
equal, and that there are trajectories yielding equal and opposite
dissipation productions whose starting points are also related by
$M^{R}$. These conditions can be expressed as:

\begin{equation}
f(M^{R}\widetilde{\Gamma},0)=f(\widetilde{\Gamma},0),
\label{eq:distcond}
\end{equation}
and

\begin{equation}
\omega_{t}(\widetilde{\Gamma})=-\omega_{t}(M^{R}\widetilde{\Gamma}{}_{t}).\label{eq:dispprodcond}
\end{equation}
$M^{R}$ is a transformation that can be more general than the map
$M^{T}$ used in Sec. \ref{sec:Methods}. $\widetilde{\Gamma}{}_{t}=S_{t}\widetilde{\Gamma}$
is the configuration to which $\widetilde{\Gamma}$ evolves after
time $t$. Provided that these two conditions hold, the derivation
of the ESFT proceeds thus. The pdf of dissipation production is

\begin{equation}
P(\omega)=\int d\widetilde{\Gamma}f(\widetilde{\Gamma},0)\delta(\omega_{t}(\widetilde{\Gamma})-\omega),
\end{equation}
and we use the definition of $\omega_{t}$ from Sec. \ref{sec:Methods}
to write

\begin{align}
P(\omega) & =\int d\widetilde{\Gamma}f(\widetilde{\Gamma},0)e^{\omega_{t}(\widetilde{\Gamma})}\frac{f(\widetilde{\Gamma}_{t},0)}{f(\widetilde{\Gamma},0)}\delta(\omega_{t}(\widetilde{\Gamma})-\omega)\nonumber \\
 & =e^{\omega}\int d\widetilde{\Gamma}f(\widetilde{\Gamma}_{t},0)\delta(\omega_{t}(\widetilde{\Gamma})-\omega).
\end{align}
Now we use the condition given in Eq. (\ref{eq:dispprodcond}) to
give

\begin{equation}
P(\omega)=e^{\omega}\int d\widetilde{\Gamma}_{t}f(\widetilde{\Gamma}_{t},0)\delta(-\omega_{t}(M^{R}\widetilde{\Gamma}_{t})-\omega).
\end{equation}
Finally, Eq. (\ref{eq:distcond}), and a transformation of the integration
measure, give

\begin{align}
P(\omega) & =e^{\omega}\int d(M^{R}\widetilde{\Gamma}_{t})f(M^{R}\widetilde{\Gamma}_{t},0)\delta(\omega_{t}(M^{R}\widetilde{\Gamma}_{t})+\omega)\nonumber \\
 & =e^{\omega}P(-\omega).
\end{align}
This is the ESFT.  Proofs in the literature employ a transformation
that time-inverts the evolved state, namely an $M^{R}$ given by $M^{T}$,
but the result can clearly hold in more general circumstances. With
$M^{R}=M^{T}$ we can employ the identity $M^{T}S_{t}^{*}M^{T}S_{t}=\mathbb{I}$
to show that condition (\ref{eq:dispprodcond}) follows from (\ref{eq:distcond}),
as long as the protocol of the dynamics is symmetric over the interval:
$S_{t}^{*}=S_{t}$. The ESFT then arises in circumstances where this
holds \emph{and }the initial pdf is symmetric in the time-reversal
operation. 

\begin{figure}
\includegraphics[scale=0.7]{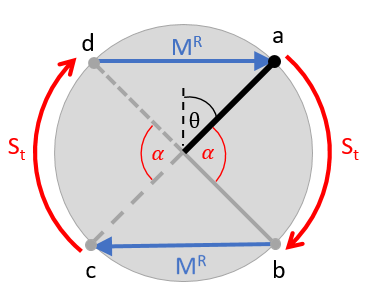}\caption{Cross-section of Bloch sphere, looking down the rotation axis of the
mapping, geometrically illustrating that $M^{R}S_{t}M^{R}S_{t}=\mathbb{I}$.
To see this, consider a point \emph{a }on the surface of the Bloch
sphere, with its position specified by angle $\theta$. Application
of evolution operator $S_{t}$ will rotate this point to $b$: $\theta\rightarrow\theta+\alpha$.
Then, $M^{R}$ will map \emph{b }to \emph{c}:\emph{ $\theta+\alpha\rightarrow2\pi-(\theta+\alpha)$.
}Applying $S_{t}$ again sends \emph{c }to \emph{d}:\emph{ }$2\pi-(\theta+\alpha)\rightarrow2\pi-(\theta+\alpha)+\alpha=2\pi-\theta$.
A final application of $M^{R}$ returns \emph{d }to \emph{a.}}
\label{fig:transformations}
\end{figure}

We anticipate that the ESFT emerges in more general circumstances
if a relation $M^{R}S_{t}M^{R}S_{t}=\mathbb{I}$ holds. This places
a requirement on the properties of $M^{R}$: in our system it must
be a reflection in the plane containing the axis of rotation of transformation
$S_{t}$ that governs the dynamics of the evolution. The requirement
$f(M^{R}\widetilde{\Gamma},0)=f(\widetilde{\Gamma},0)$ further enforces
a more stringent that $M^{R}$ is also a reflection in the plane of
symmetry of the pdf. To see this, we can associate the operations
as rotations about the axis and reflections in the plane as illustrated
in Figure \ref{fig:transformations}. 

Thus, in situations where the rotation axis representing the evolution
$S_{t}$ lies in a plane of symmetry of the pdf of the initial state
of the system, we shall observe an ESFT.

{} 

\subsubsection{Symmetry in Time Evolution into Future and Past\label{subsec:timesym}}

Provided that the conditions for obtaining an ESFT are met, specifically
that $M^{R}S_{t}M^{R}S_{t}=\mathbb{I}$ and $f(M^{R}\widetilde{\Gamma},0)=f(\widetilde{\Gamma},0)$,
we can show that the mean dissipation production is the same for evolution
into the past and the future. Starting from the mean dissipation production
for evolution into the past:

\begin{equation}
\langle\omega_{-t}\rangle=\int f(\widetilde{\Gamma},0)\mbox{ln}\frac{f(\widetilde{\Gamma},0)}{f(S_{-t}\widetilde{\Gamma},0)}d\widetilde{\Gamma},
\end{equation}
we recast as

\begin{equation}
\langle\omega_{-t}\rangle=\int f(M^{R}\widetilde{\Gamma},0)\mbox{ln}\frac{f(M^{R}\widetilde{\Gamma},0)}{f(S_{-t}M^{R}\widetilde{\Gamma},0)}dM^{R}\widetilde{\Gamma},
\end{equation}
and apply $S_{-t}M^{R}=M^{R}S_{t}$ and Eq. (\ref{eq:distcond}):

\begin{align}
\langle\omega_{-t}\rangle & =\int f(M^{R}\widetilde{\Gamma},0)\mbox{ln}\frac{f(M^{R}\widetilde{\Gamma},0)}{f(M^{R}S_{t}\widetilde{\Gamma},0)}dM^{R}\widetilde{\Gamma}\nonumber \\
 & =\int f(\widetilde{\Gamma},0)\mbox{ln}\frac{f(\widetilde{\Gamma},0)}{f(S_{t}\widetilde{\Gamma},0)}d\widetilde{\Gamma},\label{eq:16}
\end{align}
which is the mean dissipation production for forward evolution, $\langle\omega_{t}\rangle$.
However, in situations in which the ESFT is violated, we do not expect
this result to hold, the implication being that the initial ensemble
will exhibit different mean dissipation productions into the past
and the future; a time asymmetry of behaviour.

\section{Results}

In order to demonstrate the different possible behaviours of the dissipation
production, we restrict ourselves to considering simple evolutions
represented by rotations
\begin{equation}
S_{jkt}({\rm \hat{\boldsymbol{n}}},t)=\begin{cases}
\cos^{2}\frac{t}{2}+\sin^{2}\frac{t}{2}(2\hat{n}_{j}^{2}-1), & \textrm{if }j=k\\
2\hat{n}_{j}\hat{n}_{k}\sin^{2}\frac{t}{2}-\varepsilon_{jkl}\hat{n}_{l}\sin t & \textrm{if }j\neq k
\end{cases}
\end{equation}
where $\varepsilon_{jkl}$ is the Levi-Civita symbol, the rotation
angle is equal to the elapsed time, \emph{t}, and $(n_{x},n_{y},n_{z})$
is the rotation axis. 

\begin{figure}
\subfloat[Case 1a: pdf with rotational symmetry about the $z$-axis.]{\centering{}\includegraphics[width=0.5\columnwidth]{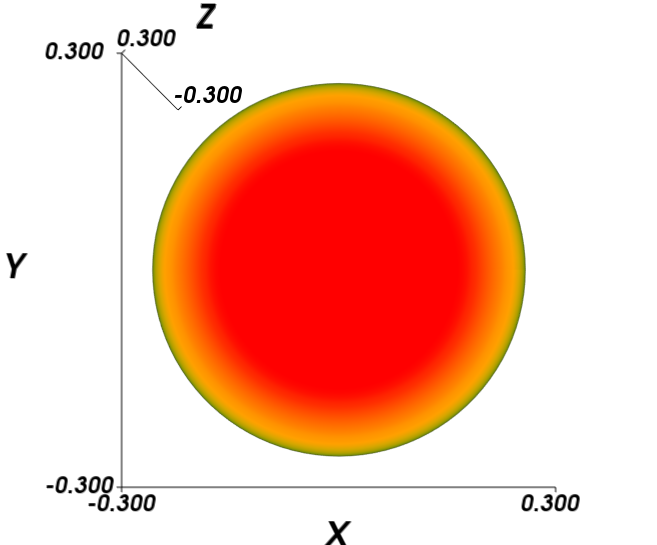}}\hfill{}\subfloat[Case 1b: pdf with rotational symmetry about $(1/\sqrt{2},1/\sqrt{2},0)$]{\centering{}\includegraphics[width=0.5\columnwidth]{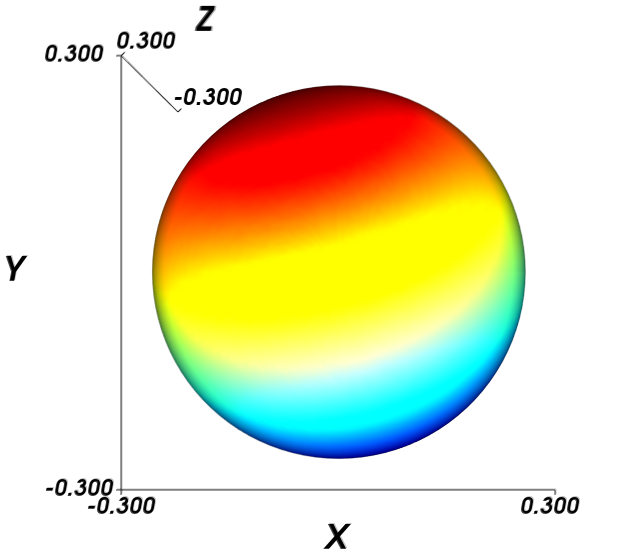}}

\subfloat[Case 2a: pdf symmetric in the $xz$-plane]{\centering{}\includegraphics[width=0.5\columnwidth]{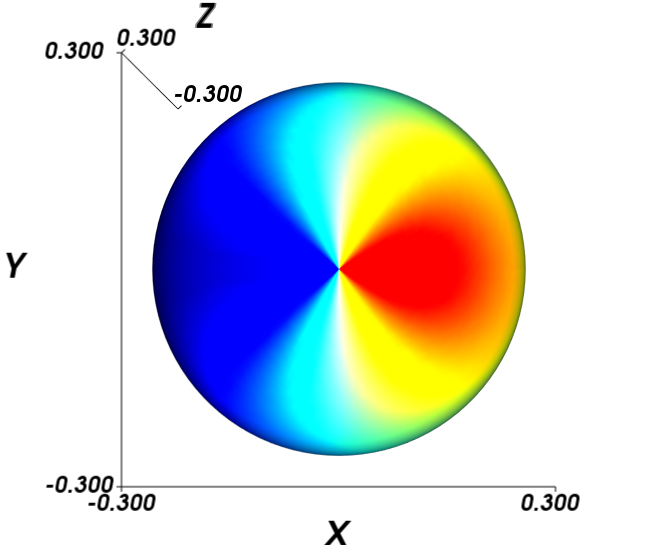}}\hfill{}\subfloat[Case 2b: pdf symmetric in plane passing through the $z$-axis]{\centering{}\includegraphics[width=0.5\columnwidth]{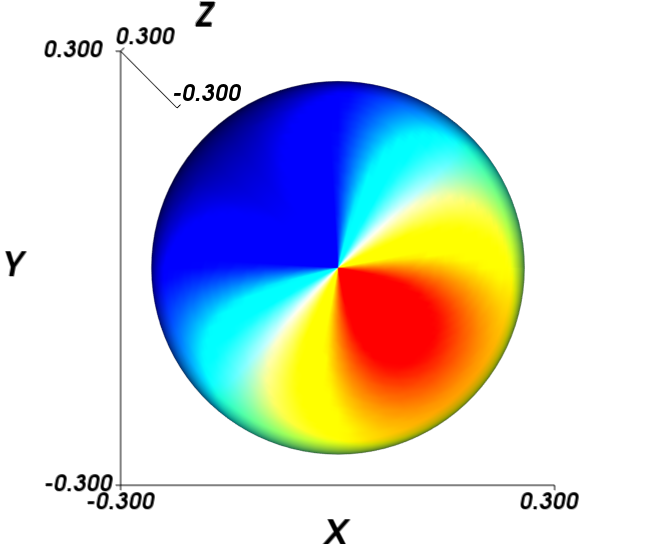}}\caption{Colour denotes the magnitude of pdf (Cases 1 and 2) at points on the
Bloch sphere, viewed from the positive $z$ direction: red is large,
blue is small.}
\label{fig:pdfs}
\end{figure}

\begin{figure}
\includegraphics[scale=0.35]{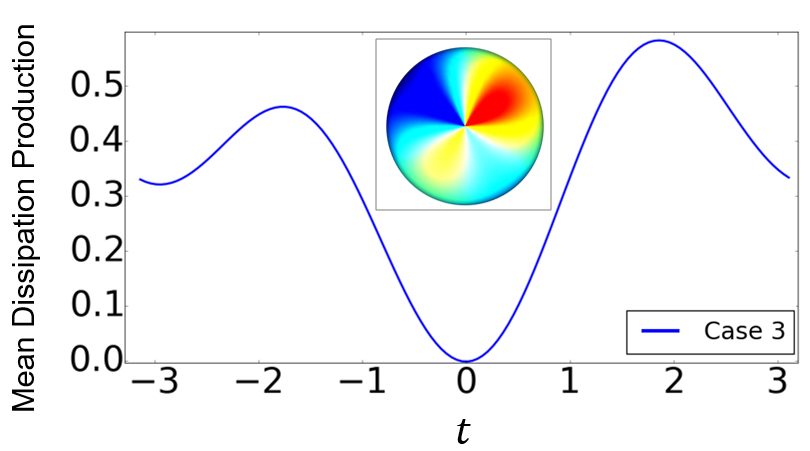}\caption{$\langle\omega_{t}\rangle$ for rotations of duration \emph{t} about
the $z$-axis in Case 3, for which the pdf is shown as an in inset.
This pdf does \emph{not} have a plane of symmetry passing through
the evolution rotation axis and an asymmetry in $\langle\omega_{t}\rangle$
for evolution into the future and the past is a consequence.}
\label{fig:asym}
\end{figure}

We consider five probability density functions over the Bloch sphere:
\begin{description}
\item [{Case~1a}] $f(\theta,\phi,t=0)=(4\pi)^{-1}(1+z)$ which is rotationally
symmetric about the $z$-axis.
\item [{Case~1b}] $f(\theta,\phi,0)=(4\pi)^{-1}(1+z\cos\beta+(u_{x}y-u_{y}x)\sin\beta+(u_{x}x+u_{y}y+u_{z}z)(1-\cos\beta)$)
with $\beta=\pi/3$, where $(u_{x},u_{y},u_{z})=(1/\sqrt{2},1/\sqrt{2},0)$
is the axis about which the pdf has rotational symmetry.
\item [{Case~2a}] $f(\theta,\phi,0)=(4\pi)^{-1}(1+\cos\theta)(1+\cos\phi),$
which is symmetric with respect to the transformation $\phi\rightarrow-\phi$
and hence has mirror symmetry in the $xz$-plane. 
\item [{Case~2b}] $f(\theta,\phi,0)=(4\pi)^{-1}(1+\cos\theta)(1+\cos(\phi+\pi/4)),$
which is \emph{not }symmetric with respect to the transformation $\phi\rightarrow-\phi$
but does have a plane of symmetry which passes through the $z$-axis. 
\item [{Case~3}] $f(\theta,\phi,0)=(8\pi)^{-1}(1+\cos\theta)(2+\cos\phi+\sin2\phi),$
which is \emph{not }symmetric with respect to the transformation $\phi\rightarrow-\phi$
and has no planes of symmetry.
\end{description}
Cases 1 and 2 are illustrated in Figure \ref{fig:pdfs} while the
fully asymmetric Case 3 is shown in Figure \ref{fig:asym}.

\begin{figure}
\includegraphics[width=1\columnwidth]{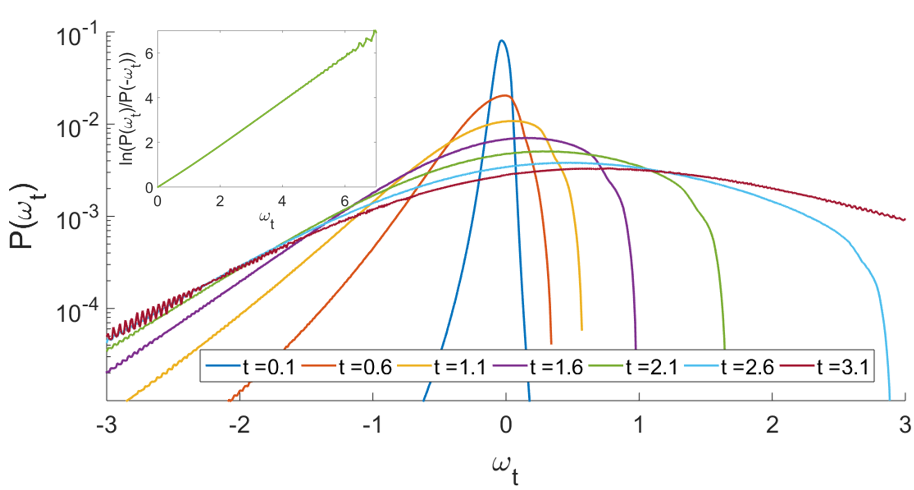}\caption{Dissipation production for Case 1a, depicted as pdfs for various elapsed
times (rotations about the $x$-axis). The inset confirms satisfaction
of an ESFT for $t=2.1$.  }
\label{fig:distributions}
\end{figure}

In Figure \ref{fig:distributions}, we show the pdfs of the dissipation
production, $\omega_{t}$, for Case 1a, using rotations about the
$x$-axis through various angles to represent the transformation $S_{t}$
at various times. The shape of the pdf in Case 1a depends on the elapsed
time. These pdfs can be used to compute the logarithm of the ratio
of probabilities of equal and opposite values of $\omega_{t}$. If
a plot of this quantity against $\omega_{t}$ gives a straight line
with unit gradient, then an ESFT holds. Since Case 1a has rotational
symmetry about the $z$-axis, any axis of rotation defining $S_{t}$
(which needn't be restricted to the Cartesian axes) will lie in a
plane of symmetry of the pdf, meeting the requirements for an ESFT
described earlier. Although the axis of rotational symmetry for Case
1b is $(1/\sqrt{2},1/\sqrt{2},0)$ rather than the $z$-axis, the
planes of symmetry through this axis nevertheless lead to the observation
of an ESFT regardless of the choice of rotation $S_{t}$. The inset
in Figure \ref{fig:distributions} demonstrates an ESFT for an example
rotation of 2.1 radians about the $x$-axis (i.e. $t=2.1$). 

\begin{figure}
\includegraphics[width=1\columnwidth]{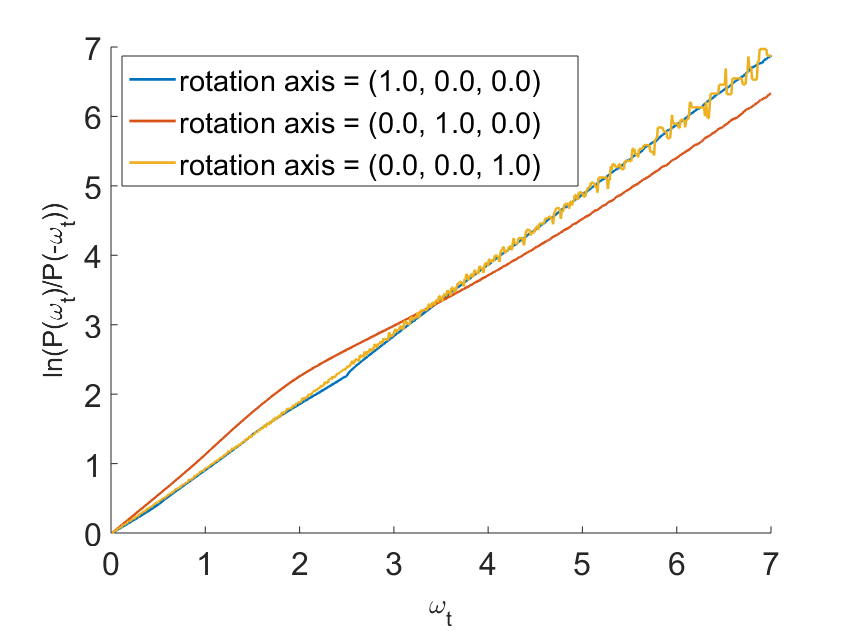}

\caption{The ESFT states that a pdf of dissipation production should satisfy
$\ln\left[P(\omega_{t})/P(-\omega_{t})\right]=\omega_{t}$. Case 2a
violates the ESFT for evolution consisting of a rotation about the
$y$-axis, which does not lie in the plane of symmetry of the pdf,
while it is upheld for rotations about the other Cartesian axes, through
which the plane of symmetry does pass. All rotations are through $2\pi/3$.}
\label{fig:flucssym} 
\end{figure}

To confirm the conditions required to obtain an ESFT, we consider
the more complicated Case 2a, which produces more structure in the
associated pdfs. Considering an evolution consisting of a rotation
angle of $2\pi/3$ about each of the Cartesian axes, we generate Figure
\ref{fig:flucssym}, which indicates, as expected, that an ESFT holds
for rotations about axes which lie in the $xz$-plane, i.e. the plane
of symmetry of the pdf on the Bloch sphere.

We also consider Case 2b which is a rotated version of the pdf in
Case 2a. We again see an ESFT holds for evolution corresponding to
rotation about the $z$-axis and its violation for rotation about
the $y$-axis, but in contrast with Case 2a, an ESFT does \emph{not
}hold for rotation about the $x$-axis as this axis does not lie in
the plane of symmetry of the Bloch sphere pdf in Case 2b.

\begin{figure}
\includegraphics[width=1\columnwidth]{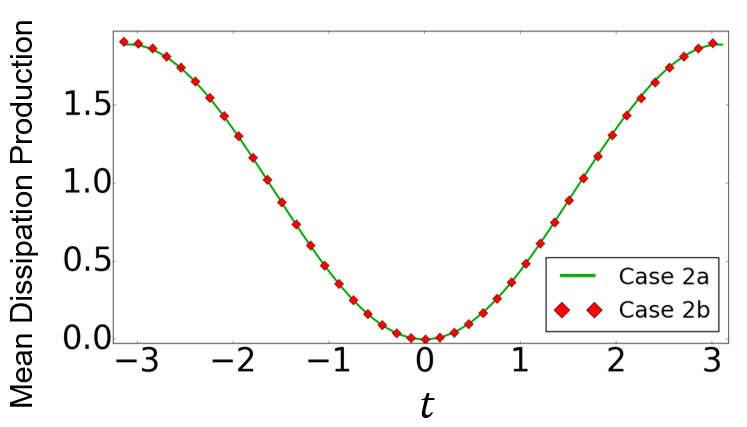}\caption{Comparison of $\langle\omega_{t}\rangle$ for rotations of duration\emph{
t} about the $z$-axis in Cases 2a and 2b. As both pdfs have a plane
of symmetry passing through the $z$-axis, $\langle\omega_{t}\rangle$
is symmetric about $t=0$, meaning that there is identical mean dissipation
production for evolution into the future and the past. Furthermore,
$\langle\omega_{t}\rangle$ is the same for both Cases. }
\label{fig:zaxisintdisp}
\end{figure}

Whenever we observe an ESFT, we also expect time-symmetric behaviour
for the mean dissipation production for evolution into the future
and the past, as explained in Sec. \ref{subsec:timesym}. Figure
\ref{fig:zaxisintdisp}, depicting $\langle\omega_{t}\rangle$ in
Cases 2a and 2b for rotations about the $z$-axis, shows this. Note
that it is non-negative as the rotation angle changes from 0 to $2\pi$.
Small angles of rotation give a small mean dissipation production,
as in these instances there is little difference between the two configurations
being compared. The non-negativity is universal and independent of
choice of axis or Bloch sphere pdf, but the time-symmetric behaviour
only accompanies situations which satisfy an ESFT.

We verify this by assessing Bloch sphere pdf Case 4, which has no
planes of symmetry, and hence cannot give an ESFT, regardless of evolution
rotation axis. Figure \ref{fig:asym} demonstrates the associated
time-asymmetry in mean dissipation production for an example rotation
about the $z$-axis.

These considerations allow us to identify initial ensembles which
will exhibit time-asymmetric average behaviour even under a reversible
unitary evolution. 

To summarise the examined Bloch sphere pdfs and rotation processes,
Table \ref{tab:table} details which combinations lead to the emergence
of an ESFT. The results confirm that an ESFT depends on the relationship
between the chosen rotation axis and the symmetry of the pdf on the
Bloch sphere; specifically when the rotation axis lies in a plane
of symmetry of the pdf, the ESFT is upheld. 

\begin{table}
\centering{}\medskip{}
\begin{tabular}{|c||c|c|c|c|}
\hline 
Axis & Cases 1a and 1b & Case 2a & Case 2b & Case 3\tabularnewline
\hline 
\hline 
$x$ & yes & yes & no & no\tabularnewline
\hline 
$y$ & yes & no & no & no\tabularnewline
\hline 
$z$ & yes & yes & yes & no\tabularnewline
\hline 
\end{tabular}\caption{Fulfillment of an ESFT for assorted cases of Bloch sphere pdfs when
configurations are rotated by the dynamics $S_{t}$ about the Cartesian
axes.}
\label{tab:table}
\end{table}

\section{Conclusion}

In quantum systems undergoing deterministic evolution, the ensemble
that represents our uncertainty with regard to the initial state vector
can be used to specify the likelihood of following a particular trajectory
and its reversed counterpart. This allows us to test for irreversibility
of behaviour in the form of the failure of \textit{obversibility}
(a property distinct from, but closely related to \emph{reversibility}),
which we quantify with \emph{dissipation production}. This first study
of dissipation production in a quantum system extends the use of the
concepts beyond the classical realm previously considered \citep{Ford}.
In particular, we are able to determine whether a system with a given
pdf describing an initial ensemble will exhibit time-asymmetry in
average behaviour under a particular process.

We have studied a simple two-level system, which is nevertheless sufficient
to demonstrate the use of dissipation production for situations in
which entropy production is an inappropriate irreversibility measure.
Our principal aim has been to identify conditions under which dissipation
production satisfies an Evans-Searles Fluctuation Theorem (ESFT),
in which case it follows that it evolves on average into the past
in the same way as into the future. The mapping of states on the Bloch
sphere after a given time interval under deterministic dynamics can
be represented by a rotation about a certain axis, and the criterion
for the validity of the ESFT is that this axis should lie in a plane
of symmetry of the pdf describing the initial ensemble. It is also
straightforward to demonstrate that the average dissipation production
can never be negative, which makes it a measure of irreversibility. 

Obversibility is distinct from reversibility. The latter is upheld
here owing to the deterministic unitary dynamics employed: the system
is isolated. Reversibility is (essentially) the property that the
effects of carrying out a process, given an ensemble of initial system
configurations, can be undone by inverting velocities, carrying out
a reverse process, and then inverting velocities again. Obversibility
is the property that the effects of a process and a reverse process,
preceded by an inversion of velocities in the latter case, but starting
from a given ensemble, are statistically identical \citep{Ford}.

Dissipation production is a consequence of a failure of obversibility
and plays a role that is similar to, but distinct from the entropy
production that arises from a failure of reversibility. In a nonequilibrium
stationary state, dissipation production and entropy production are
synonymous, but not in general situations. We have therefore broadened
our understanding of quantities that might characterise the arrow
of development in time. Furthermore, we have been able to demonstrate
that the time-asymmetric nature of this arrow for a closed quantum
system can arise from certain asymmetries in the pdf over the ensemble
of initial states. As dissipation production depends only on the classical
probabilities of starting from particular states, even when considering
quantum systems, it emphasises the distinction between classical uncertainty
due to a lack of knowledge regarding, for example, the initial state,
and quantum uncertainty due to a lack of predictability with regard
to the outcomes of measurement; we currently exclude the latter from
our considerations.

We anticipate that the methods described here are readily applicable
to larger systems, such as two qubits, since the tools required to
calculate dissipation production (namely pdfs of the system configuration
and appropriate reversal and evolution operators) can be readily defined.
Although we have illustrated our examples with the Bloch sphere, this
is not an essential component. For a general complex system it is
likely that the ESFT will be upheld under very special circumstances,
but these will always include situations where the pdf describing
the initial ensemble obeys time-reversal symmetry and the protocol
of dynamics is time-symmetric about its midpoint, as envisaged by
Evans \citep{evans1993probability}.  

In summary we have confirmed that the failure of obversibility can
be used as an indicator of irreversibility in a closed system, where
mechanical reversibility is respected. Such a failure can also be
used to characterise irreversibility in systems that are \emph{open}
to the environment, by way of measurement and/or thermalisation. The
dynamics of such systems are mechanically irreversible, and we would
compute dissipation production merely by inserting appropriate transition
probabilities into Eq. (\ref{eq:fulldisprod}). The relationship between
reversibility and obversibility needs to be further developed, giving
consideration to the role of initial conditions in generating subsequent
time-asymmetry of the irreversibility measures. Exploring dissipation
production and obversibility in open quantum situations is hence an
avenue for further research.

\begin{acknowledgments}
This work was supported by the U.K. Engineering and Physical Sciences
Research Council through the Centre for Doctoral Training in Delivering
Quantum Technologies at UCL, grant number 1489394.
\end{acknowledgments}

\bibliographystyle{apsrev4-1}
\bibliography{obvbiblio}

\begin{thebibliography}{17}%
\makeatletter
\providecommand \@ifxundefined [1]{%
 \@ifx{#1\undefined}
}%
\providecommand \@ifnum [1]{%
 \ifnum #1\expandafter \@firstoftwo
 \else \expandafter \@secondoftwo
 \fi
}%
\providecommand \@ifx [1]{%
 \ifx #1\expandafter \@firstoftwo
 \else \expandafter \@secondoftwo
 \fi
}%
\providecommand \natexlab [1]{#1}%
\providecommand \enquote  [1]{``#1''}%
\providecommand \bibnamefont  [1]{#1}%
\providecommand \bibfnamefont [1]{#1}%
\providecommand \citenamefont [1]{#1}%
\providecommand \href@noop [0]{\@secondoftwo}%
\providecommand \href [0]{\begingroup \@sanitize@url \@href}%
\providecommand \@href[1]{\@@startlink{#1}\@@href}%
\providecommand \@@href[1]{\endgroup#1\@@endlink}%
\providecommand \@sanitize@url [0]{\catcode `\\12\catcode `\$12\catcode
  `\&12\catcode `\#12\catcode `\^12\catcode `\_12\catcode `\%12\relax}%
\providecommand \@@startlink[1]{}%
\providecommand \@@endlink[0]{}%
\providecommand \url  [0]{\begingroup\@sanitize@url \@url }%
\providecommand \@url [1]{\endgroup\@href {#1}{\urlprefix }}%
\providecommand \urlprefix  [0]{URL }%
\providecommand \Eprint [0]{\href }%
\providecommand \doibase [0]{http://dx.doi.org/}%
\providecommand \selectlanguage [0]{\@gobble}%
\providecommand \bibinfo  [0]{\@secondoftwo}%
\providecommand \bibfield  [0]{\@secondoftwo}%
\providecommand \translation [1]{[#1]}%
\providecommand \BibitemOpen [0]{}%
\providecommand \bibitemStop [0]{}%
\providecommand \bibitemNoStop [0]{.\EOS\space}%
\providecommand \EOS [0]{\spacefactor3000\relax}%
\providecommand \BibitemShut  [1]{\csname bibitem#1\endcsname}%
\let\auto@bib@innerbib\@empty
\bibitem [{\citenamefont {Kreuzer}(1981)}]{kreuzer1981nonequilibrium}%
  \BibitemOpen
  \bibfield  {author} {\bibinfo {author} {\bibfnamefont {H.~J.}\ \bibnamefont
  {Kreuzer}},\ }\href@noop {} {\bibfield  {journal} {\bibinfo  {journal}
  {Oxford and New York, Clarendon Press, 1981. 455 p.}\ }\textbf {\bibinfo
  {volume} {1}} (\bibinfo {year} {1981})}\BibitemShut {NoStop}%
\bibitem [{\citenamefont {Lebowitz}(1993)}]{lebowitz1993boltzmann}%
  \BibitemOpen
  \bibfield  {author} {\bibinfo {author} {\bibfnamefont {J.~L.}\ \bibnamefont
  {Lebowitz}},\ }\href@noop {} {\bibfield  {journal} {\bibinfo  {journal}
  {Physics Today}\ }\textbf {\bibinfo {volume} {46}},\ \bibinfo {pages} {32}
  (\bibinfo {year} {1993})}\BibitemShut {NoStop}%
\bibitem [{\citenamefont {Albert}(2009)}]{albert2009time}%
  \BibitemOpen
  \bibfield  {author} {\bibinfo {author} {\bibfnamefont {D.~Z.}\ \bibnamefont
  {Albert}},\ }\href@noop {} {\emph {\bibinfo {title} {Time and Chance}}}\
  (\bibinfo  {publisher} {Harvard University Press},\ \bibinfo {year}
  {2009})\BibitemShut {NoStop}%
\bibitem [{\citenamefont {Tolman}\ and\ \citenamefont
  {Fine}(1948)}]{tolman1948irreversible}%
  \BibitemOpen
  \bibfield  {author} {\bibinfo {author} {\bibfnamefont {R.~C.}\ \bibnamefont
  {Tolman}}\ and\ \bibinfo {author} {\bibfnamefont {P.~C.}\ \bibnamefont
  {Fine}},\ }\href@noop {} {\bibfield  {journal} {\bibinfo  {journal} {Reviews
  of Modern Physics}\ }\textbf {\bibinfo {volume} {20}},\ \bibinfo {pages} {51}
  (\bibinfo {year} {1948})}\BibitemShut {NoStop}%
\bibitem [{\citenamefont {Eckmann}\ and\ \citenamefont
  {Ruelle}(1985)}]{eckmann1985ergodic}%
  \BibitemOpen
  \bibfield  {author} {\bibinfo {author} {\bibfnamefont {J.-P.}\ \bibnamefont
  {Eckmann}}\ and\ \bibinfo {author} {\bibfnamefont {D.}~\bibnamefont
  {Ruelle}},\ }\href@noop {} {\bibfield  {journal} {\bibinfo  {journal}
  {Reviews of Modern Physics}\ }\textbf {\bibinfo {volume} {57}},\ \bibinfo
  {pages} {617} (\bibinfo {year} {1985})}\BibitemShut {NoStop}%
\bibitem [{\citenamefont {Deffner}\ and\ \citenamefont
  {Lutz}(2011)}]{deffner2011nonequilibrium}%
  \BibitemOpen
  \bibfield  {author} {\bibinfo {author} {\bibfnamefont {S.}~\bibnamefont
  {Deffner}}\ and\ \bibinfo {author} {\bibfnamefont {E.}~\bibnamefont {Lutz}},\
  }\href@noop {} {\bibfield  {journal} {\bibinfo  {journal} {Physical Review
  Letters}\ }\textbf {\bibinfo {volume} {107}},\ \bibinfo {pages} {140404}
  (\bibinfo {year} {2011})}\BibitemShut {NoStop}%
\bibitem [{\citenamefont {Horowitz}\ and\ \citenamefont
  {Parrondo}(2013)}]{horowitz2013entropy}%
  \BibitemOpen
  \bibfield  {author} {\bibinfo {author} {\bibfnamefont {J.~M.}\ \bibnamefont
  {Horowitz}}\ and\ \bibinfo {author} {\bibfnamefont {J.~M.}\ \bibnamefont
  {Parrondo}},\ }\href@noop {} {\bibfield  {journal} {\bibinfo  {journal} {New
  Journal of Physics}\ }\textbf {\bibinfo {volume} {15}},\ \bibinfo {pages}
  {085028} (\bibinfo {year} {2013})}\BibitemShut {NoStop}%
\bibitem [{\citenamefont {Elouard}\ \emph {et~al.}(2017)\citenamefont
  {Elouard}, \citenamefont {Bernardes}, \citenamefont {Carvalho}, \citenamefont
  {Santos},\ and\ \citenamefont {Auff{\`e}ves}}]{elouard2017probing}%
  \BibitemOpen
  \bibfield  {author} {\bibinfo {author} {\bibfnamefont {C.}~\bibnamefont
  {Elouard}}, \bibinfo {author} {\bibfnamefont {N.~K.}\ \bibnamefont
  {Bernardes}}, \bibinfo {author} {\bibfnamefont {A.~R.}\ \bibnamefont
  {Carvalho}}, \bibinfo {author} {\bibfnamefont {M.~F.}\ \bibnamefont
  {Santos}}, \ and\ \bibinfo {author} {\bibfnamefont {A.}~\bibnamefont
  {Auff{\`e}ves}},\ }\href@noop {} {\bibfield  {journal} {\bibinfo  {journal}
  {New Journal of Physics}\ } (\bibinfo {year} {2017})}\BibitemShut {NoStop}%
\bibitem [{\citenamefont {Ford}(2015)}]{Ford}%
  \BibitemOpen
  \bibfield  {author} {\bibinfo {author} {\bibfnamefont {I.~J.}\ \bibnamefont
  {Ford}},\ }\href {http://stacks.iop.org/1367-2630/17/i=7/a=075017} {\bibfield
   {journal} {\bibinfo  {journal} {New Journal of Physics}\ }\textbf {\bibinfo
  {volume} {17}},\ \bibinfo {pages} {075017} (\bibinfo {year}
  {2015})}\BibitemShut {NoStop}%
\bibitem [{\citenamefont {Evans}\ and\ \citenamefont
  {Searles}(2002)}]{evans2002fluc}%
  \BibitemOpen
  \bibfield  {author} {\bibinfo {author} {\bibfnamefont {D.~J.}\ \bibnamefont
  {Evans}}\ and\ \bibinfo {author} {\bibfnamefont {D.~J.}\ \bibnamefont
  {Searles}},\ }\href@noop {} {\bibfield  {journal} {\bibinfo  {journal}
  {Advances in Physics}\ }\textbf {\bibinfo {volume} {51}},\ \bibinfo {pages}
  {1529} (\bibinfo {year} {2002})}\BibitemShut {NoStop}%
\bibitem [{\citenamefont {Evans}\ \emph {et~al.}(1993)\citenamefont {Evans},
  \citenamefont {Cohen},\ and\ \citenamefont {Morriss}}]{evans1993probability}%
  \BibitemOpen
  \bibfield  {author} {\bibinfo {author} {\bibfnamefont {D.~J.}\ \bibnamefont
  {Evans}}, \bibinfo {author} {\bibfnamefont {E.}~\bibnamefont {Cohen}}, \ and\
  \bibinfo {author} {\bibfnamefont {G.}~\bibnamefont {Morriss}},\ }\href@noop
  {} {\bibfield  {journal} {\bibinfo  {journal} {Physical Review Letters}\
  }\textbf {\bibinfo {volume} {71}},\ \bibinfo {pages} {2401} (\bibinfo {year}
  {1993})}\BibitemShut {NoStop}%
\bibitem [{\citenamefont {Haake}(2013)}]{haake2013quantum}%
  \BibitemOpen
  \bibfield  {author} {\bibinfo {author} {\bibfnamefont {F.}~\bibnamefont
  {Haake}},\ }\href@noop {} {\emph {\bibinfo {title} {Quantum Signatures of
  Chaos}}}\ (\bibinfo  {publisher} {Springer Science \& Business Media},\
  \bibinfo {year} {2013})\BibitemShut {NoStop}%
\bibitem [{\citenamefont {Nielsen}\ and\ \citenamefont
  {Chuang}(2002)}]{nielsen2002quantum}%
  \BibitemOpen
  \bibfield  {author} {\bibinfo {author} {\bibfnamefont {M.~A.}\ \bibnamefont
  {Nielsen}}\ and\ \bibinfo {author} {\bibfnamefont {I.}~\bibnamefont
  {Chuang}},\ }\href@noop {} {\emph {\bibinfo {title} {Quantum Computation and
  Quantum Information}}}\ (\bibinfo  {publisher} {AAPT},\ \bibinfo {year}
  {2002})\BibitemShut {NoStop}%
\bibitem [{\citenamefont {Reid}\ \emph {et~al.}(2012)\citenamefont {Reid},
  \citenamefont {Evans},\ and\ \citenamefont
  {Searles}}]{reid2012communication}%
  \BibitemOpen
  \bibfield  {author} {\bibinfo {author} {\bibfnamefont {J.~C.}\ \bibnamefont
  {Reid}}, \bibinfo {author} {\bibfnamefont {D.~J.}\ \bibnamefont {Evans}}, \
  and\ \bibinfo {author} {\bibfnamefont {D.~J.}\ \bibnamefont {Searles}},\
  }\href@noop {} {\bibfield  {journal} {\bibinfo  {journal} {The Journal of
  Chemical Physics}\ }\textbf {\bibinfo {volume} {136}},\ \bibinfo {pages}
  {021101} (\bibinfo {year} {2012})}\BibitemShut {NoStop}%
\bibitem [{\citenamefont {Kullback}\ and\ \citenamefont
  {Leibler}(1951)}]{kullback1951information}%
  \BibitemOpen
  \bibfield  {author} {\bibinfo {author} {\bibfnamefont {S.}~\bibnamefont
  {Kullback}}\ and\ \bibinfo {author} {\bibfnamefont {R.~A.}\ \bibnamefont
  {Leibler}},\ }\href@noop {} {\bibfield  {journal} {\bibinfo  {journal} {The
  Annals of Mathematical Statistics}\ }\textbf {\bibinfo {volume} {22}},\
  \bibinfo {pages} {79} (\bibinfo {year} {1951})}\BibitemShut {NoStop}%
\bibitem [{\citenamefont {Harris}\ and\ \citenamefont
  {Sch{\"u}tz}(2007)}]{harris2007fluctuation}%
  \BibitemOpen
  \bibfield  {author} {\bibinfo {author} {\bibfnamefont {R.}~\bibnamefont
  {Harris}}\ and\ \bibinfo {author} {\bibfnamefont {G.}~\bibnamefont
  {Sch{\"u}tz}},\ }\href@noop {} {\bibfield  {journal} {\bibinfo  {journal}
  {Journal of Statistical Mechanics: Theory and Experiment}\ ,\ \bibinfo
  {pages} {P07020}} (\bibinfo {year} {2007})}\BibitemShut {NoStop}%
\bibitem [{\citenamefont {Crooks}(1999)}]{PhysRevE.60.2721}%
  \BibitemOpen
  \bibfield  {author} {\bibinfo {author} {\bibfnamefont {G.~E.}\ \bibnamefont
  {Crooks}},\ }\href {\doibase 10.1103/PhysRevE.60.2721} {\bibfield  {journal}
  {\bibinfo  {journal} {Phys. Rev. E}\ }\textbf {\bibinfo {volume} {60}},\
  \bibinfo {pages} {2721} (\bibinfo {year} {1999})}\BibitemShut {NoStop}%
\end{thebibliography}%

\end{document}